\documentclass[12pt]{article}

\usepackage[top=2cm, bottom=2cm, left=2cm, right=2cm]{geometry}
\usepackage{amsmath}
\usepackage{amssymb}
\usepackage{mathrsfs}
\usepackage{graphicx}
\usepackage{lineno}

\usepackage{hyperref} 
\usepackage{xcolor}   

\title{The large obliquity of Saturn explained by the fast migration of Titan}

\author{Melaine Saillenfest$^1$ \and Giacomo Lari$^2$ \and Gwena{\"e}l Bou{\'e}$^1$}

\date{\color{blue}This is the author's manuscript of the letter published in Nature Astronomy on 18-01-2021, \emph{Nat Astron 5, 345--349 (2021)} \url{https://doi.org/10.1038/s41550-020-01284-x}}

\begin{document}
\maketitle

{\footnotesize

   \noindent $^1$ IMCCE, Observatoire de Paris, PSL Research University, CNRS, Sorbonne Universit\'e, Universit\'e de Lille, 75014 Paris, France
   
   \noindent $^2$ Department of Mathematics, University of Pisa, Largo Bruno Pontecorvo 5, 56127 Pisa, Italy
}
\bigskip


{\bf

The obliquity of a planet is the tilt between its equator and its orbital plane. Giant planets are expected to form with near-zero obliquities \cite{Ward-Hamilton_2004,Rogoszinski-Hamilton_2020a}. After its formation, some dynamical mechanism must therefore have tilted Saturn up to its current obliquity of $26.7^\circ$. This event is traditionally thought to have happened more than $4$~Gyrs ago during the late planetary migration \cite{Malhotra_1993,Tsiganis-etal_2005,Nesvorny-Morbidelli_2012} because of the crossing of a resonance between the spin-axis precession of Saturn and the nodal orbital precession mode of Neptune \cite{Hamilton-Ward_2004}. Here, we show that the fast tidal migration of Titan measured by \cite{Lainey-etal_2020} is incompatible with this scenario, and that it offers a new explanation for Saturn's current obliquity. A significant migration of Titan would prevent any early resonance, invalidating previous constraints on the late planetary migration set by the tilting of Saturn \cite{Boue-etal_2009,Brasser-Lee_2015,Vokrouhlicky-Nesvorny_2015}. We propose instead that the resonance was encountered recently, about $1$~Gyr ago, forcing Saturn's obliquity to increase from a small value (possibly less than $3^\circ$), up to its current state. This scenario suggests that Saturn's normalised polar moment of inertia lies between $0.224$ and $0.237$. Our findings bring out a new paradigm for the spin-axis evolution of Saturn, Jupiter \cite{Saillenfest-etal_2020}, and possibly giant exoplanets in multiple systems, whereby obliquities are not settled once for all, but continuously evolve as a result of the migration of their satellites.

}

We investigate whether the spin-axis dynamics of Saturn could have been influenced by the migration of its satellites. The torque applied by the sun on Saturn's equatorial bulge produces a precession of its spin axis with a mean frequency $\dot{\psi} = \alpha\,(1-e^2)^{-3/2}\cos\varepsilon$, where $e$ is the eccentricity of Saturn, $\varepsilon$ is its obliquity, and $\alpha$ is called its precession constant \cite{Laskar-Robutel_1993}. The value of $\alpha$ incorporates the extra torques produced by Saturn's satellites \cite{Ward-Hamilton_2004,Ward_1975}. It also depends on Saturn's orbital and physical parameters, among which all are well known except Saturn's normalised polar moment of inertia $\lambda$. Estimates found in the literature broadly range in $\lambda\in[0.200,0.240]$ but they are model-dependent and a proper uncertainty range is hard to define (see Methods). Exploring this whole interval gives a current precession constant $\alpha_0$ ranging from $0.747$ to $0.894''/$yr.

\begin{figure}
   \centering
   \includegraphics[width=0.7\textwidth]{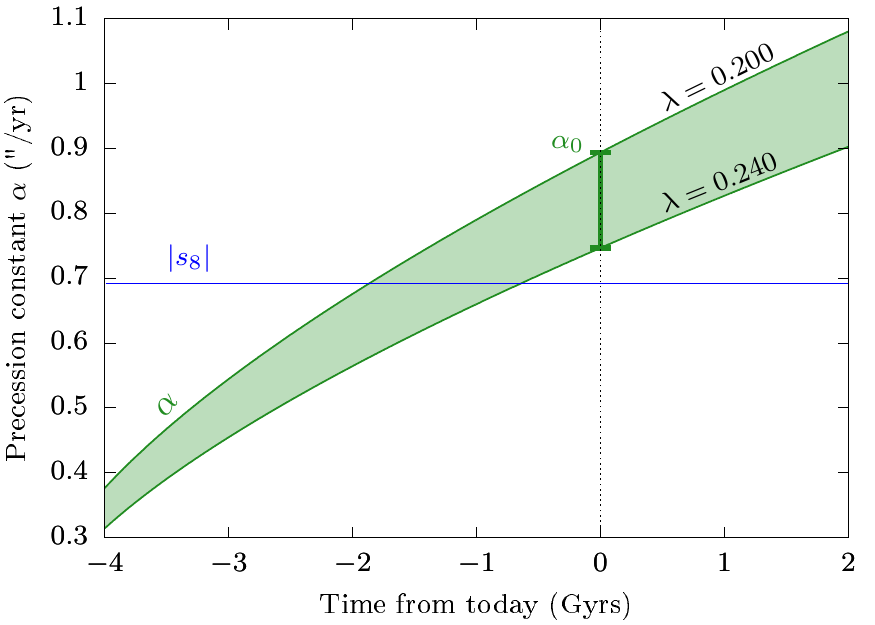}
   \caption{Evolution of the effective precession constant $\alpha$ of Saturn due to the tidal migration of Titan. The green error bar shows the current value $\alpha_0$ obtained by considering Saturn's normalised polar moment of inertia $\lambda$ lying in $[0.200,0.240]$. For a given value of $\lambda$ (see labels), $\alpha$ is propagated through time using Titan's nominal migration law given by \cite{Lainey-etal_2020}. The blue line shows the value of the nodal precession mode of Neptune $s_8$; it is nearly constant since the end of the late planetary migration \cite{Laskar_1990}, and was higher before. The spin-axis precession rate of Saturn scales as $\alpha\cos\varepsilon$, where $\varepsilon$ is its obliquity.}
   \label{fig:alphaevol}
\end{figure}

We take into account the migration of Saturn's satellites by changing their contributions to~$\alpha$. The measurements of \cite{Lainey-etal_2020} unveiled a fast migration for Titan, and a similar tidal timescale $t_\text{tide}=a/\dot{a}$ for all six satellites of Saturn studied. Both these findings support the ``resonance locking'' tidal theory of \cite{Fuller-etal_2016}. Using this theory, \cite{Lainey-etal_2020} obtained an approximate formula for the long-term trend of Titan's semi-major axis $a$ over time, in which Titan's tidal timescale evolves in the long run with a current value $t_\text{tide}\approx 13$~Gyrs, in accordance with observations. We use this formula in order to estimate the variations of $\alpha$ over time, and since Titan is by far the satellite that contributes most, we include other satellites through a slight increase of Titan's mass (see Methods). Titan's mean semi-major axis is today $a=20.275$ radii of Saturn ($R_\mathrm{eq}$), but \cite{Lainey-etal_2020} suggest that during the late planetary migration, more than $4$~Gyrs ago, it was smaller than $10$\,$R_\mathrm{eq}$. As illustrated in Fig.~\ref{fig:alphaevol}, this variation translates into a large-scale drift of Saturn's precession constant $\alpha$. Therefore, contrary to previous assumptions \cite{Boue-etal_2009,Brasser-Lee_2015,Vokrouhlicky-Nesvorny_2015}, we find that $\alpha$ cannot be considered as roughly constant over time.

A planet undergoes a secular spin-orbit resonance when there is a commensurability between the precession frequency of its spin axis and the frequency of one or several harmonics appearing in its orbital precession \cite{Laskar-Robutel_1993,Millholland-Laughlin_2019,Saillenfest-etal_2019a}. The orbital motion of the giant planets is now very steady \cite{Laskar_1990}, such that the nodal orbital precession mode of Neptune $s_8\approx -0.6919''/$yr is almost constant. In contrast, Fig.~\ref{fig:alphaevol} displays values of $\alpha$ ranging between $0.314$ and $0.376''$/yr after the end of the late planetary migration. These values are much smaller than $s_8$ or any other orbital frequency of the planets \cite{Laskar_1990}, preventing Saturn from an early encounter with a low-order secular spin-orbit resonance.

This result is not specific to the migration law used for Titan; it arises for any realistic law compatible with the measurements of \cite{Lainey-etal_2020} -- unless we are currently witnessing a transient episode of exceptionally fast migration for Titan and coincidentally similar $t_\text{tide}$ for most major satellites of Saturn, which appears unlikely. Moreover, since the planets were more tightly packed during their late migration, their orbital precession frequencies were all higher than today \cite{Brasser-Lee_2015}, whereas $\alpha$ was not much affected by the slight change of Saturn's semi-major axis. This stretches even further the mismatch between both kinds of precession frequencies. The ring of Saturn, even if it was more massive than today \cite{Charnoz-etal_2018}, did not contribute significantly, and a massive circumplanetary disc could only have produced a small obliquity kick (see Methods). Consequently, Saturn's obliquity remained essentially unaltered at that time. Provided that Titan did migrate substantially before today (i.e. over a few $R_\mathrm{eq}$ or more), this result would invalidate previous constraints on the late planetary migration set by the tilting of Saturn; the two events are probably unrelated.

\begin{figure}
   \centering
   \includegraphics[width=0.53\textwidth]{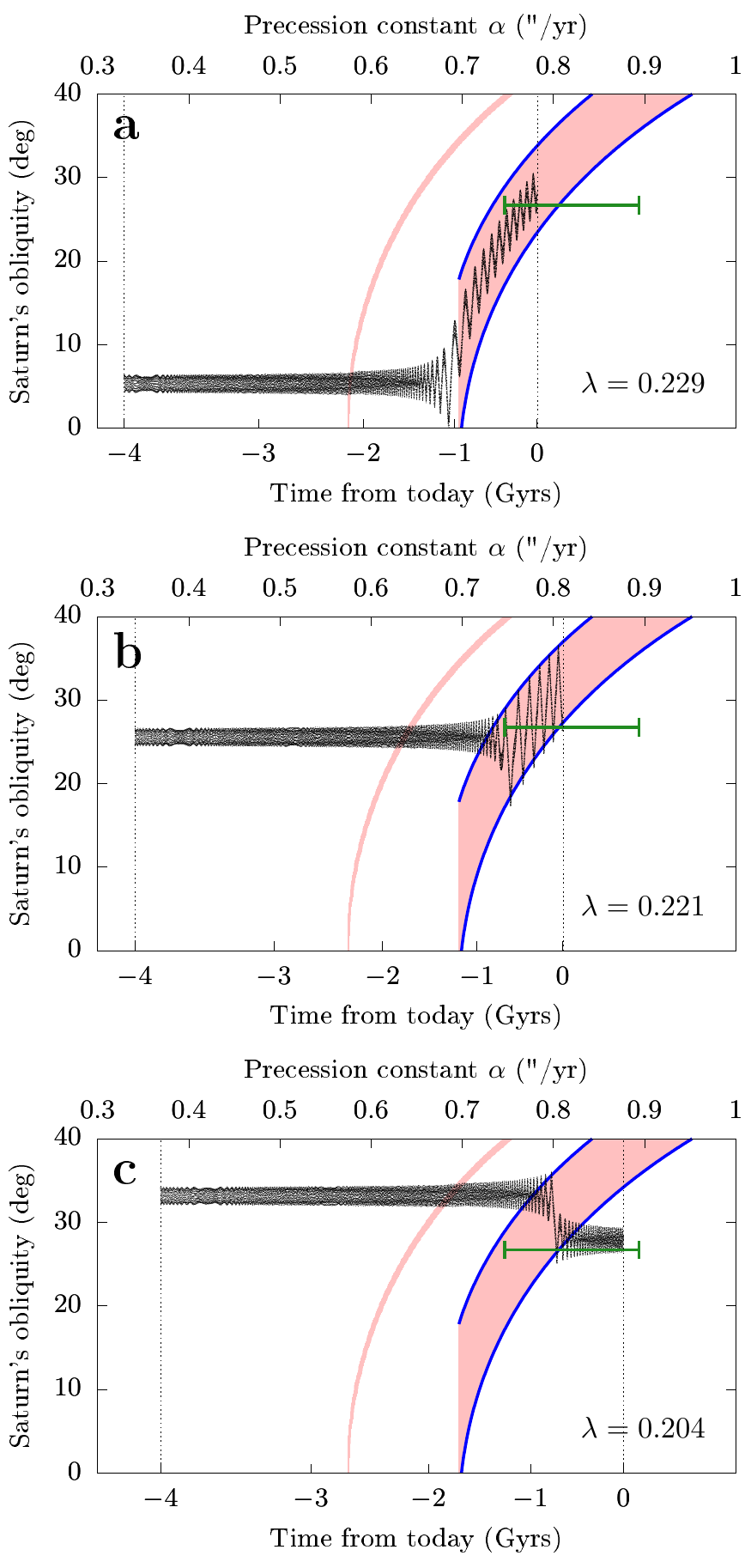}
   \caption{Examples of possible obliquity evolution for Saturn. Each graph shows a $4$-Gyr numerical trajectory of Saturn (black dots) computed for $\alpha$ evolving as in Fig.~\ref{fig:alphaevol}. Saturn lies today along the green error bar (same as Fig.~\ref{fig:alphaevol} at $t=0$), and the value of $\lambda$ used is written on each panel. The pink areas show all first-order secular spin-orbit resonances \cite{Saillenfest-etal_2019a}. The large right area is the resonance with $s_8$ and the left thin area is the resonance with $g_7-g_8+s_7$ \cite{Laskar_1990}. The separatrices of the $s_8$ resonance are highlighted in blue. If we assume that Saturn's primordial precession angle $\psi$ is a random number uniformly distributed between $0$ and $2\pi$, trajectories encountering the separatrix would have a given probability of being captured (as in panel \textbf{b}) or crossing over the resonance (as in panel \textbf{c}). A given value of $\lambda$, however, corresponds to one and only one dynamical pathway, like those shown in these examples.}
   \label{fig:timeevol}
\end{figure}

Instead, the fast migration of Titan implies that Saturn reached the resonance with $s_8$ much more recently, as a result of the increase of $\alpha$. In order to explore possible trajectories of Saturn's spin axis, we propagate it numerically backwards in time for $\lambda$ finely sampled in its exploration interval. We use the same setting as \cite{Saillenfest-etal_2020}: the orbit of Saturn evolves according to the full series of \cite{Laskar_1990} and the precession constant $\alpha$ is taken as a slowly-varying parameter. Numerical experiments show that the system is not chaotic enough to produce noticeable numerical irreversibility, such that backward propagating does reveal the only dynamical pathway corresponding to a given state of Saturn today. We limit our propagations to $4$~Gyrs in the past, that is, confidently after the late planetary migration. As illustrated in Fig.~\ref{fig:timeevol}, we find that the resonance with $s_8$ was reached about $1$~Gyr ago. Before this event, the obliquity of Saturn oscillated with a small amplitude around a fixed value. The crossing of the thin resonance with $g_7-g_8+s_7$ hardly affected it at all. During the late planetary migration that occurred before our $4$-Gyr timespan, the planetary precession frequencies were all higher than today, resulting in an even steadier obliquity value.

\begin{figure}
   \centering
   \includegraphics[width=0.6\textwidth]{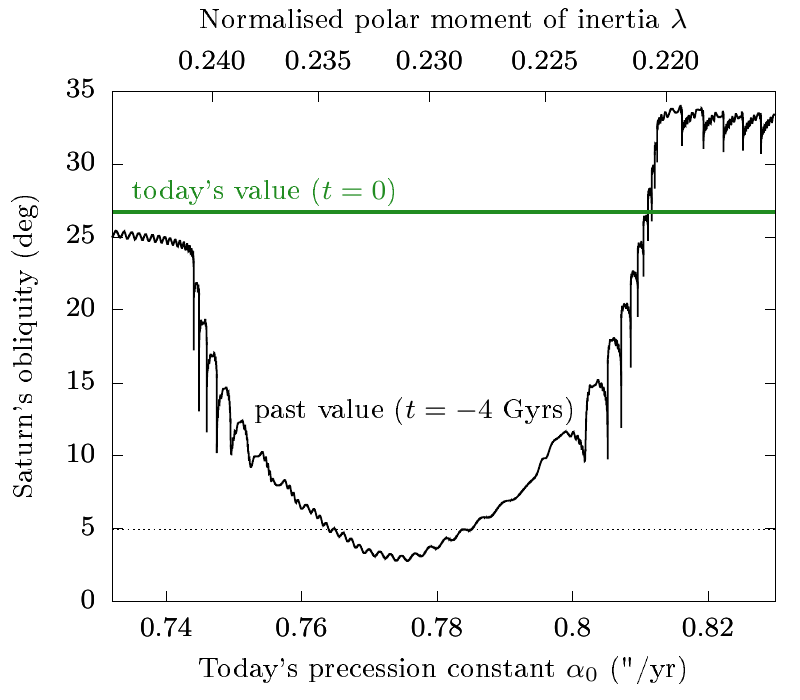}
   \caption{Past obliquity of Saturn as a function of its current precession constant $\alpha_0$, for Titan migrating as in Fig.~\ref{fig:alphaevol}. Each point of the black curve is obtained from a numerical simulation (see Fig.~\ref{fig:timeevol} for examples). The green line shows Saturn's current obliquity and the black dotted line highlights the $5^\circ$ level.}
   \label{fig:obminnom}
\end{figure}

\begin{figure}
   \centering
   \includegraphics[width=0.8\textwidth]{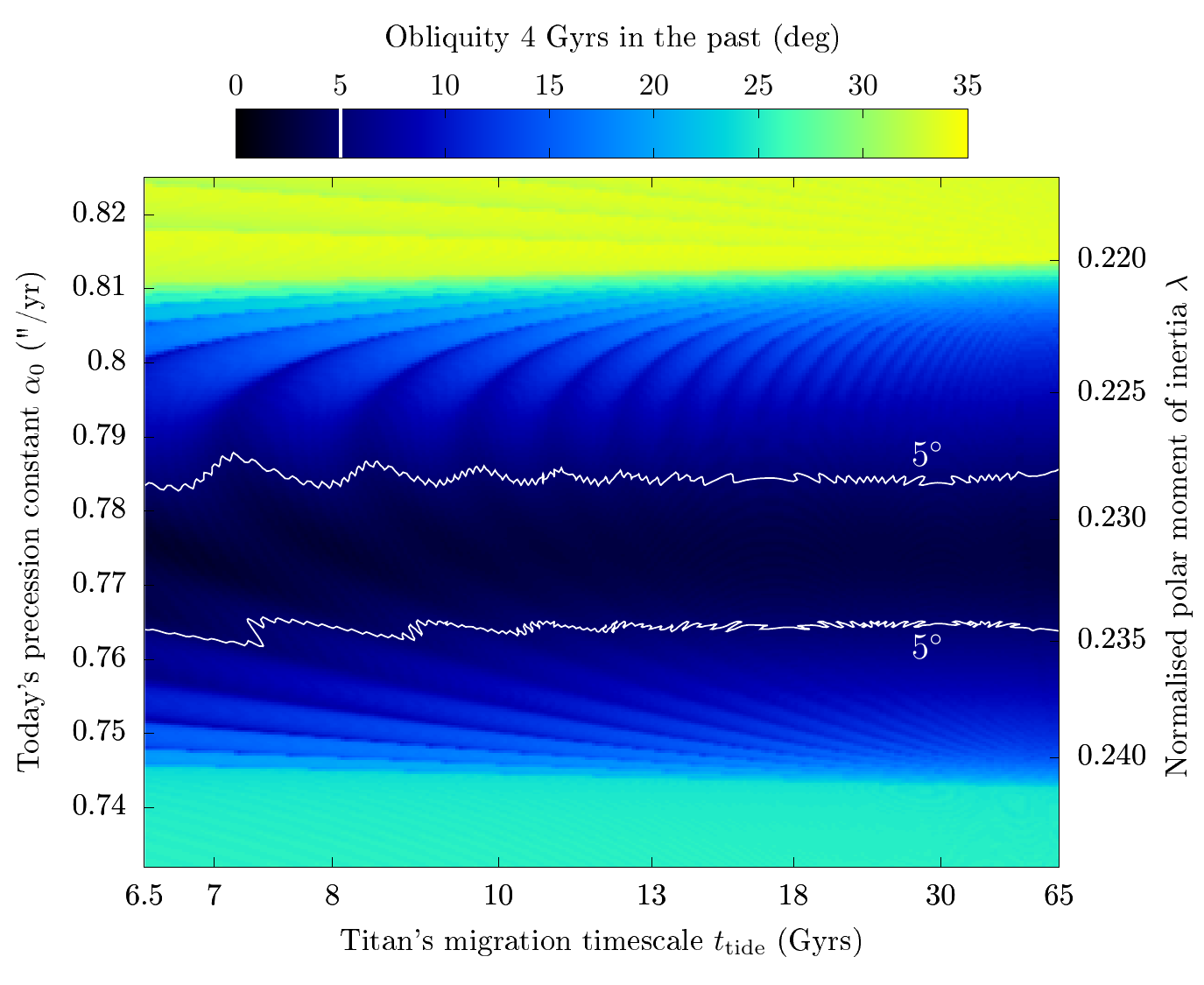}
   \caption{Past obliquity of Saturn for different migration histories of Titan. On the horizontal axis, $t_\text{tide}=a/\dot{a}$ is Titan's current migration timescale. The horizontal axis is not linear but scaled as $1/t_\text{tide}$. Between the white curves, Saturn's past obliquity is smaller than $5^\circ$. The curve of Fig.~\ref{fig:obminnom} corresponds here to a vertical slice at $t_\mathrm{tide}\approx 13$~Gyrs. Assuming that $t_\mathrm{tide}=6.5$~Gyrs instead, the resonance with $s_8$ would have been encountered about $0.4$~Gyrs ago, whereas assuming that $t_\mathrm{tide}=65$~Gyrs, it would have been encountered about $2.8$~Gyrs ago.}
   \label{fig:obminzoom}
\end{figure}

Distinct dynamical pathways can have been followed by Saturn according to its current precession constant $\alpha_0$, whose value depends on Saturn's polar moment of inertia $\lambda$. Values of $\lambda$ ranging between about $0.224$ and $0.237$ imply that Saturn currently oscillates with a small amplitude inside the resonance. In this case, when Saturn's precession constant went over $\alpha\approx 0.696''/$yr, the separatrices of the resonance appeared around its trajectory, leading to a $100\%$-sure capture (panel \textbf{a} of Fig.~\ref{fig:timeevol}). This is what we expect if Saturn reached the resonance with a small obliquity \cite{Ward-Hamilton_2004}. Other values of $\lambda$ imply either that Saturn was recently captured with an already large obliquity (panel \textbf{b}), or that Saturn went past the resonance through its hyperbolic point (panel \textbf{c}). Figure~\ref{fig:obminnom} shows a summary of Saturn's past obliquity values obtained. The U-shaped portion of the curve corresponds to values of $\alpha_0$ putting today Saturn inside the resonance with $s_8$. If Saturn is currently out of resonance, previous studies -- even though they relied on a different scenario -- also reported the impossibility (or negligible probability) of producing Saturn's current obliquity from an initially low value  \cite{Hamilton-Ward_2004,Boue-etal_2009,Brasser-Lee_2015,Vokrouhlicky-Nesvorny_2015}. The bulk of successful runs in previous studies correspond to values of $\alpha_0$ putting today Saturn in the vicinity of the resonance centre, which is also what we observe here in the context of Titan's tidal migration. Figure~\ref{fig:obminzoom} features a minimum past obliquity of Saturn equal to about $2.5^\circ$, reached for $\alpha_0\approx 0.775''/$yr (i.e. $\lambda\approx 0.231$).

The current orientation of Saturn's spin axis implies that if $\alpha$ evolved in a purely adiabatic way (i.e. very slowly with respect to the obliquity oscillations), then Saturn's primordial obliquity cannot have been exactly zero \cite{Ward-Hamilton_2004}. However, the migration rate of Titan measured by \cite{Lainey-etal_2020} is so fast that it departs somewhat from adiabaticity, and one could expect that different migration laws would lead to different outcomes. The influence of Titan's migration law can be investigated by varying its tidal timescale $t_\mathrm{tide}=a/\dot{a}$. The nominal law of \cite{Lainey-etal_2020} corresponds today to $t_\text{tide}\approx 13$~Gyrs, and radio-science experiments yield values ranging between $9.3$ and $13.2$~Gyrs at the $3\sigma$ uncertainty level. Since Titan's fast migration may have been episodic \cite{Fuller-etal_2016}, we extend our analysis beyond this interval. Our results are summarised in Fig.~\ref{fig:obminzoom}. In spite of shallow wavy patterns, Fig.~\ref{fig:obminzoom} has a very horizontal-like structure, showing that Titan's migration is an almost adiabatic process. Therefore, Saturn's spin-axis dynamics mostly depend on the range over which Titan migrates and not much on its migration rate. For instance, the results obtained for $t_\mathrm{tide}=65$~Gyrs are similar to those obtained in a scenario in which Titan remained unmoved for billions of years and then started to migrate at its observed rate during the last $1.5$~Gyrs only. This example shows that our results are valid for a large variety of migration histories compatible with the measurements of \cite{Lainey-etal_2020}, and even if Titan's migration was sporadic. The minimum past obliquity of Saturn is firmly below $5^\circ$, in an almost unvaried range of $\alpha_0$ values. The remaining slightly non-zero excess of obliquity can be explained by many different phenomena, like the light bombardment phase occurring at the end of Saturn's formation, or an abrupt resonance crossing stemming from the dissipation of the protoplanetary disc \cite{Millholland-Batygin_2019}. The tidal migration of Titan is therefore able to explain the current large obliquity of Saturn starting from a small, presumably primordial, value.

This scenario is however not unique. On the one hand, our current knowledge of the tidal resonance-locking mechanism of \cite{Fuller-etal_2016} leaves the possibility that we are now observing a very fast migration of Titan which otherwise remained nearly zero over \emph{all} Titan's history. This extreme possibility appears yet unlikely, since the resonance-locking mechanism seemingly triggered for all six moons of Saturn observed by \cite{Lainey-etal_2020}, favouring a continuous rather than sporadic migration. On the other hand, the possibility remains that Saturn acquired its large obliquity via an early impact \cite{Safronov_1966} and was then not much affected by the recent resonance crossing with $s_8$ (panels \textbf{b} and \textbf{c} of Fig.~\ref{fig:timeevol}). Ruling out this possibility would require a better knowledge of Saturn's polar moment of inertia, hopefully obtained in a model-independent way.

The mechanism presented in this letter is generic and also applies to Jupiter, even though Jupiter did not have enough time yet to be substantially tilted \cite{Saillenfest-etal_2020}. Giant exoplanets could harbour massive satellites as well \cite{Kipping_2014}. If a fast satellite migration is indeed the rule \cite{Lainey-etal_2020}, the obliquities of such planets are vulnerable to dramatic increase via the same mechanism. The existence of secular spin-orbit resonances only requires a precessing orbit, for instance due to other planetary companions \cite{Millholland-Laughlin_2019}. The number of strong harmonics in the precession spectrum of a planet is larger for planetary systems featuring substantial eccentricities and inclinations. The various orbital architectures observed in exoplanetary systems should therefore allow for much more diverse secular spin-orbit resonances than for the giant planets of the solar system, which have today rather cold orbits \cite{Kreyche-etal_2020}. The mechanism presented in this letter could therefore greatly affect the obliquity distribution of exoplanets.

\section*{Methods}
\paragraph{Long-term dynamics of Saturn's spin axis} The long-term spin-axis dynamics of an oblate planet is ruled by the Hamiltonian function
\begin{equation}\label{eq:Hinit}
   \begin{aligned}
      \mathcal{H}(X,-\psi,t) &= -\frac{\alpha}{2}\frac{X^2}{\big(1-e(t)^2\big)^{3/2}} \\
      &- \sqrt{1-X^2}\big(\mathcal{A}(t)\sin\psi + \mathcal{B}(t)\cos\psi\big) \\
      &+ 2X\mathcal{C}(t),
   \end{aligned}
\end{equation}
where the conjugate coordinates are $X=\cos\varepsilon$ (cosine of obliquity) and $-\psi$ (minus the precession angle) \cite{Laskar-Robutel_1993}. The quantity $\alpha$ is called the ``precession constant''. Equation~\eqref{eq:Hinit} explicitly depends on time $t$ through the orbital eccentricity $e$ of the planet and through the functions
\begin{equation}
   \left\{
   \begin{aligned}
      \mathcal{A}(t) &= \frac{2\big(\dot{q}+p\,\mathcal{C}(t)\big)}{\sqrt{1-p^2-q^2}}\,, \\
      \mathcal{B}(t) &= \frac{2\big(\dot{p}-q\,\mathcal{C}(t)\big)}{\sqrt{1-p^2-q^2}} \,,\\
   \end{aligned}
   \right.
   \quad
   \text{and}
   \quad
   \mathcal{C}(t) = q\dot{p}-p\dot{q}\,.
\end{equation}
In these expressions, $q=\eta\cos\Omega$ and $p=\eta\sin\Omega$, where $\eta\equiv\sin(I/2)$, and $I$ and $\Omega$ are the orbital inclination and the longitude of ascending node of the planet, respectively. If the planet has migrating satellites, its precession constant $\alpha$ is a slowly varying function of time.

\paragraph{Saturn's precession constant} The precession constant of a planet can be written
\begin{equation}\label{eq:alp}
   \alpha = \frac{3}{2}\frac{\mathcal{G}m_\odot}{\omega a_\mathrm{S}^3}\frac{J_2 + q}{\lambda + \ell}\,,
\end{equation}
where $q$ and $\ell$ are the satellites' contributions \cite{Ward-Hamilton_2004,Ward_1975}. Here, $\mathcal{G}m_\odot$ is the sun's gravitational parameter, $\omega$ is Saturn's spin rate, $a_\mathrm{S}$ its semi-major axis, $J_2$ its second zonal gravity coefficient, and $\lambda$ is its polar moment of inertia $C$ normalised using its mass $M$ and its equatorial radius $R_\mathrm{eq}$ as $\lambda\equiv C/(MR_\mathrm{eq}^2)$.

When computing $\alpha$ through Eq.~\eqref{eq:alp}, the values of $J_2$, $\lambda$, $q$, and $\ell$ must be normalised using the same reference radius so that the final value of $\alpha$ is independent of the normalisation used. Different normalisation conventions are the source of much confusion in the literature. Here, we use quantities normalised using $R_\mathrm{eq} = 60268$~km; values taken from other articles are renormalised accordingly. The choice of $a_\mathrm{S}$ in Eq.~\eqref{eq:alp} is also critical, since it can lead to substantially different values of $\alpha$ and place today Saturn inside or outside the resonance with $s_8$. Since Eq.~\eqref{eq:alp} relates to secular theories (i.e. averaged over rotational and orbital motions), one must use secular elements as those computed by \cite{Laskar_1990}. The spin rate and current spin-axis orientation of Saturn are given by the latest IAU recommendations \cite{Archinal-etal_2018} and its $J_2$ by \cite{Iess-etal_2019}.

Saturn's normalised polar moment of inertia $\lambda$ is still poorly constrained today. Even though many different values can be found is the literature, they are all model-dependent and they generally disagree with each other. Citing \cite{Hubbard-Marley_1989}, authors often consider $\lambda=0.22$ as a central value, with a $10\%$ error meant to account for model-dependence \cite{French-etal_1993,Ward-Hamilton_2004,Boue-etal_2009}. This corresponds to an interval that is slightly larger than $\lambda\in[0.200,0.240]$. In more recent articles, it seems still unclear what are the actual uncertainties of estimates of $\lambda$, even when taking into account the \emph{Cassini} data. When reviewing values obtained through different models, specialists still stick to $\lambda=0.22$ with a $10\%$ variation \cite{Fortney-etal_2018}.

Values of $q$ and $\ell$ including the meaningful contributions from all satellites of Saturn have been computed by \cite{French-etal_1993}. Here, we use updated values of $q$ and $\ell$ obtained by taking into account the non-zero equilibrium orbital inclination (``Laplace plane'') for all satellites and by using the satellites' masses and mean orbital elements coming from modern ephemerides (JPL SAT427\footnote{\texttt{https://ssd.jpl.nasa.gov/}}). Compared to \cite{French-etal_1993}, these refined values of $q$ and $\ell$ lead to a slightly different relation between the value of Saturn's current precession constant $\alpha_0$ and the value of Saturn's normalised polar moment of inertia $\lambda$.

\paragraph{Migration of Saturn's satellites} The contributions $q$ and $\ell$ of migrating satellites vary over time, producing a drift of $\alpha$. Through the tidal theory of \cite{Fuller-etal_2016}, the semi-major-axis variations of most major satellites of Saturn are approximated by \cite{Lainey-etal_2020} as
\begin{equation}\label{eq:aTit}
   a(t) \approx a_0\left(\frac{t}{t_0}\right)^B\,,
\end{equation}
where $a_0$ is the satellite's current mean semi-major axis, $t_0$ is Saturn's current age, and $B\approx 1/3$ is a constant. The tidal timescale $t_\text{tide}=a/\dot{a}$ of the satellites can be derived from Eq.~\eqref{eq:aTit}: it evolves in the long run, with a current value $t_\text{tide}= t_0/B \approx 13$~Gyrs. 

Being so much more massive, Titan is by far the satellite that contributes most to the value of $\alpha$. The second largest contribution comes from Iapetus, but it is only $5\%$ of Titan's \cite{French-etal_1993}. Therefore, most of the drift of $\alpha$ over time is due to Titan's migration. For this reason and because of the quasi-adiabatic nature of the drift, other satellites can be modelled as a slight increase of Titan's mass, allowing us to only include Titan in Eq.~\eqref{eq:alp}. The adjusted mass of Titan is about $4\%$ larger. A more refined model taking into account the migration of all satellites (or variations of Saturn's physical parameters in Eq.~\ref{eq:alp}) would only slightly change the evolution timescale of $\alpha$ illustrated in Fig.~\ref{fig:alphaevol}, and not our conclusions. The slight increase of Titan's mass is here only meant to provide the right connection between $\lambda$ and today's value of $\alpha$, such that $\alpha(t=0) = \alpha_0$.

The effect of a past massive ring on the value of $\alpha$ can be estimated by putting an annulus of material at Saturn's fluid Roche radius in Eq.~\eqref{eq:alp}. Giving it five times the mass of Titan, as suggested by \cite{Hesselbrock-Minton_2017}, does not substantially modifies the value of $\alpha$. Our conclusions about the past small value of $\alpha$ even hold if the annulus is several tens of times more massive than Titan. Only a wide and massive long-lived circumplanetary disc could possibly have enhanced $\alpha$ enough to reach a resonance \cite{Millholland-Batygin_2019}, but its decay would then only have produced a small obliquity kick \cite{Hamilton-Ward_2004}.

\paragraph{Orbital solution of Saturn} The Hamiltonian function in Eq.~\eqref{eq:Hinit} depends on the orbit of the planet and on its temporal variations. We need a secular orbital solution that is valid over billions of years, which is well beyond the timespan covered by ephemerides. Luckily, the orbital dynamics of the giant planets of the solar system are almost integrable and excellent solutions have been developed. We use the secular solution of \cite{Laskar_1990} expanded in quasi-periodic series:
\begin{equation}\label{eq:qprep}
   \begin{aligned}
      z = e\exp(i\varpi) &= \sum_k E_k\exp(i\theta_k) \,,\\
      \zeta = \eta\exp(i\Omega) &= \sum_k S_k\exp(i\phi_k)\,,
   \end{aligned}
\end{equation}
where $\varpi$ is Saturn's longitude of perihelion. The amplitudes $E_k$ and $S_k$ are real constants, and the angles $\theta_k$ and $\phi_k$ evolve linearly over time $t$, with frequencies $\mu_k$ and $\nu_k$:
\begin{equation}
   \theta_k(t) = \mu_k\,t + \theta_k^{(0)}
   \hspace{0.5cm}\text{and}\hspace{0.5cm}
   \phi_k(t) = \nu_k\,t + \phi_k^{(0)}\,.
\end{equation}
As described by \cite{Saillenfest-etal_2020}, the complete secular orbital solution of \cite{Laskar_1990} is obtained by multiplying the normalised proper modes $z_i^\bullet$ and $\zeta_i^\bullet$ (Tables VI and VII of \cite{Laskar_1990}) by  the matrix $\tilde{S}$ corresponding to the linear part of the solution (Table V of \cite{Laskar_1990}). In the series obtained, the terms with the same combination of frequencies are then merged together, resulting in $56$ terms in eccentricity and $60$ terms in inclination, with amplitudes down to $10^{-8}$. These terms contain contributions from all the planets of the solar system. Even though frequencies associated to the terrestrial planets could vary somewhat over billions of years as a result of their chaotic dynamics \cite{Laskar_1990}, they only marginally contribute to Saturn's orbital solution and none of them takes part in the resonances shown in Fig.~\ref{fig:timeevol}. This secular orbital solution can therefore be considered valid over a billion-year timescale, since the end of the late planetary migration.

Because the orbital series of Saturn are here not restricted to the $s_8$ harmonic, our results differ to some degree from previous analytical predictions. In particular, the minimum past obliquity of Saturn is about $2.5^\circ$ for an adiabatic resonance capture (see main text). This value is smaller than the $4.5^\circ$ previously predicted \cite{Ward-Hamilton_2004}, because of extra obliquity oscillations produced by other terms of the orbital series.

\paragraph{Numerical integration} The dynamical system described by Eq.~\eqref{eq:Hinit} is propagated numerically. The precession constant $\alpha$ is made to vary according to the migration of Titan in Eq.~\eqref{eq:aTit}, and the orbit of Saturn is made to vary according to the quasi-periodic series in Eq.~\eqref{eq:qprep}. Instead of the variables $(X,\psi)$ shown in Eq.~\eqref{eq:Hinit}, numerical calculations are made with the equivalent spherical coordinates of the spin axis
\begin{equation}
   \left\{
   \begin{aligned}
      x &= \sin(\varepsilon)\cos(\psi)\,, \\
      y &= \sin(\varepsilon)\sin(\psi)\,, \\
      z &= \cos(\varepsilon)\,,
   \end{aligned}
   \right.
\end{equation}
which are not singular for a zero obliquity. Numerical integrations are performed with small enough step-size so that $\sqrt{x^2+y^2+z^2}$ is conserved to within the computer's round-off errors.

\section*{Data availability}
Data for Figs.~\ref{fig:alphaevol} to \ref{fig:obminzoom} are available as Source Data with the paper. The data supporting other findings of this study are available from the corresponding author upon reasonable request.

\section*{Code availability}
The parameters and equations of motion are fully described within the paper. All data can be reproduced using any standard implementation. The numerical integration scheme used is fully available from \cite{Rein-Spiegel_2015}.

\bibliographystyle{naturemag}
\bibliography{lettersaturnspin}

\section*{Additional information}
\paragraph{Correspondence and requests for materials} should be addressed to M.~S.

\section*{Acknowledgements}
We thank Luis Gomez Casajus for fruitful discussions and St{\'e}fan Renner for his suggestions during the writing of our manuscript. G.~L. acknowledges financial support from the Italian Space Agency (ASI) through agreement 2017-40-H.0.

\section*{Author contributions}
G.~B. brought out the original idea. G.~L. compiled the data. G.~L. and M.~S. made the computations. M.~S. wrote the article. All authors participated in supervising the whole study.

\section*{Competing interests}
The authors declare no competing financial interests.

\newpage

\end{document}